\documentclass[aps,reprint,showpacs,superscriptaddress,longbibliography]{revtex4-2} 
\usepackage{hyperref} 
\hypersetup
{
    colorlinks=true,       
    linkcolor=blue,        
    citecolor=blue,        
    urlcolor=blue           
}
\usepackage{amsmath}
\usepackage{amssymb} 
\usepackage{bm}
\usepackage{graphicx} 
\usepackage{subfigure} 
\usepackage{dcolumn} 
\usepackage{color}
\usepackage{textcomp} %
\usepackage{ulem}
\usepackage{booktabs}       
\usepackage[table]{xcolor}  
\usepackage{amssymb}        
\usepackage{graphicx}       
\usepackage{cancel}

\usepackage{pifont}

\usepackage{kotex}

\usepackage{float}

\usepackage{xcolor}

\newcommand{\rev}[1]{{\color{black} #1}}

\begin{document}
\title{Deterministic N\'eel vector switching of altermagnets via magnetic octupole torque} 
\author{Seungyun Han}  
\affiliation{Department of Physics, Pohang University of Science and Technology, Pohang 37673, Korea}
\affiliation{Center for Quantum Dynamics of Angular Momentum, Pohang University of Science and Technology, Pohang 37673, Korea}
\author{Insu Baek}  
\affiliation{Department of Physics, Pohang University of Science and Technology, Pohang 37673, Korea}
\affiliation{Center for Quantum Dynamics of Angular Momentum, Pohang University of Science and Technology, Pohang 37673, Korea}
\author{Kyoung-Whan Kim}
\email{kwkim@yonsei.ac.kr}
\affiliation{Department of Physics, Yonsei University, Seoul 03722, Korea}
\author{Hyun-Woo Lee}
\email{hwl@postech.ac.kr}
\affiliation{Department of Physics, Pohang University of Science and Technology, Pohang 37673, Korea}
\affiliation{Center for Quantum Dynamics of Angular Momentum, Pohang University of Science and Technology, Pohang 37673, Korea}
\author{Suik Cheon}
\email{enprodigy@postech.ac.kr}
\affiliation{Department of Physics, Pohang University of Science and Technology, Pohang 37673, Korea} 
\affiliation{Center for Quantum Dynamics of Angular Momentum, Pohang University of Science and Technology, Pohang 37673, Korea}
\date{\today}
\begin{abstract}
Altermagnets have recently emerged as promising materials for next-generation spintronic devices. For their device applications, realizing a single-domain configuration is essential but remains challenging. 
We theoretically consider injecting magnetic multipoles into altermagnets, which can be achieved by applying an in-plane current to 
an altermagnet/normal metal bilayer.
We demonstrate for $d$-wave altermagnets that the torque generated by the magnetic octupole injection can achieve magnetic-field-free deterministic switching of the altermagnets' N\'eel vector and transform their multidomain configurations into a single domain.
This method allows the switching in diverse altermagnets,
thereby facilitating their device applications and fundamental studies. This work also exemplifies the usefulness of magnetic multipole currents.
\end{abstract}
%
\maketitle 
%
%
%
 
{\it Introduction--} Altermagnets (AMs) represent a new class of magnetism~\cite{hayami2019, naka2019, ahn2019,yuan2020, igor2021, yuan2021, yuan2022_1, yuan2022_2} that combines characteristics of ferromagnets (FMs) and antiferromagnets (AFMs). Whereas their spins are compensated like AFMs with opposite spins on the sublattices A and B, they exhibit FM-like features such as non-relativistic spin splitting~\cite{wu2007, noda2016, ma2021, smejkal2022_1, smejkal2022_2,lee2024,han2024} and the anomalous Hall effect~\cite{smejkal2020, shao2021}. These properties lead to interesting effects such as the spin-splitter Hall effect and spin-splitter torque~\cite{gonzalez2021, smejkal2022_1, karube2022, bose2022}, which are useful for device applications. However, experimentally measured effects are significantly weaker than theoretical predictions, primarily due to the multidomain issue inherent in AMs~\cite{shao2021,liu2024, noh2025}.

To utilize the intriguing properties of AMs for device applications, it is desired to develop a method to control the N\'eel vector~\cite{han2024, zhou2025} and force AMs into a single-domain configuration. However, due to their AFM nature, achieving this goal remains elusive. Aligning the N\'eel vector using an external magnetic field is limited to AMs with net magnetization~\cite{amin2024}. Moreover, this method is not suitable from a device application perspective. Recently, a few ideas to realize electrical control were reported. One method exploits broken inversion symmetry in AMs~\cite{chen2025, song2025}. However, this method is
not applicable to centrosymmetric AMs. Another method utilizes N\'eel spin currents~\cite{shao2023, zhang2025}, which are effective only in AMs with special sublattice structures.

We explore another method that utilizes a defining difference of AMs from conventional AFMs: magnetic multipoles~\cite{han2024octupole}. Recent studies have identified magnetic multipoles as secondary order parameters in AMs. For instance, $d$-wave AMs exhibit ferroic magnetic octupole (MO) ordering, while $g$-wave AMs are characterized by magnetic triakontadipole ordering~\cite{bhowal2024,mcclarty2024, verbeek2024}. Additionally, magnetic multipoles couple to the N\'eel vector of AMs in a manner analogous to the coupling in FMs between magnetic dipoles (spin) and local magnetizations. Due to the coupling, injecting a spin current into FMs generates the spin-transfer torque (STT) or the spin-orbit torque (SOT) on the local magnetization in FMs~\cite{gambardella2011, khvalkovskiy2013, manchon2019}. Some of us recently demonstrated that injecting a magnetic multipole current into AMs similarly generates  a magnetic multipole torque on the N\'eel vector~\cite{han2024octupole} in AMs.

In this Letter, we explore the analogy to the FM case further and examine the possibility that the magnetic multipole torque may open a novel paradigm of the N\'eel vector control in AMs just as the STT or the SOT induces the deterministic switching in FMs~\cite{gambardella2011, khvalkovskiy2013, manchon2019}. As a specific example,
we examine the N\'eel vector dynamics in $d$-wave AMs induced by the MO torque (MOT). The electrically induced MOT generates staggered effective magnetic fields to the sublattices of $d$-wave AMs~\cite{han2024octupole}. We show that the MOT can switch the N\'eel vector deterministically without the assistance of an external magnetic field and transform AMs in multidomain magnetic configurations into a single magnetic domain. This mechanism does not require the net magnetization, the Dzyaloshinskii-Moriya interaction, the inversion symmetry breaking, or special sublatice structures of the AMs.

{\it Schematic--}
\begin{figure}[]
  \centering
  \includegraphics[width=8.8cm]{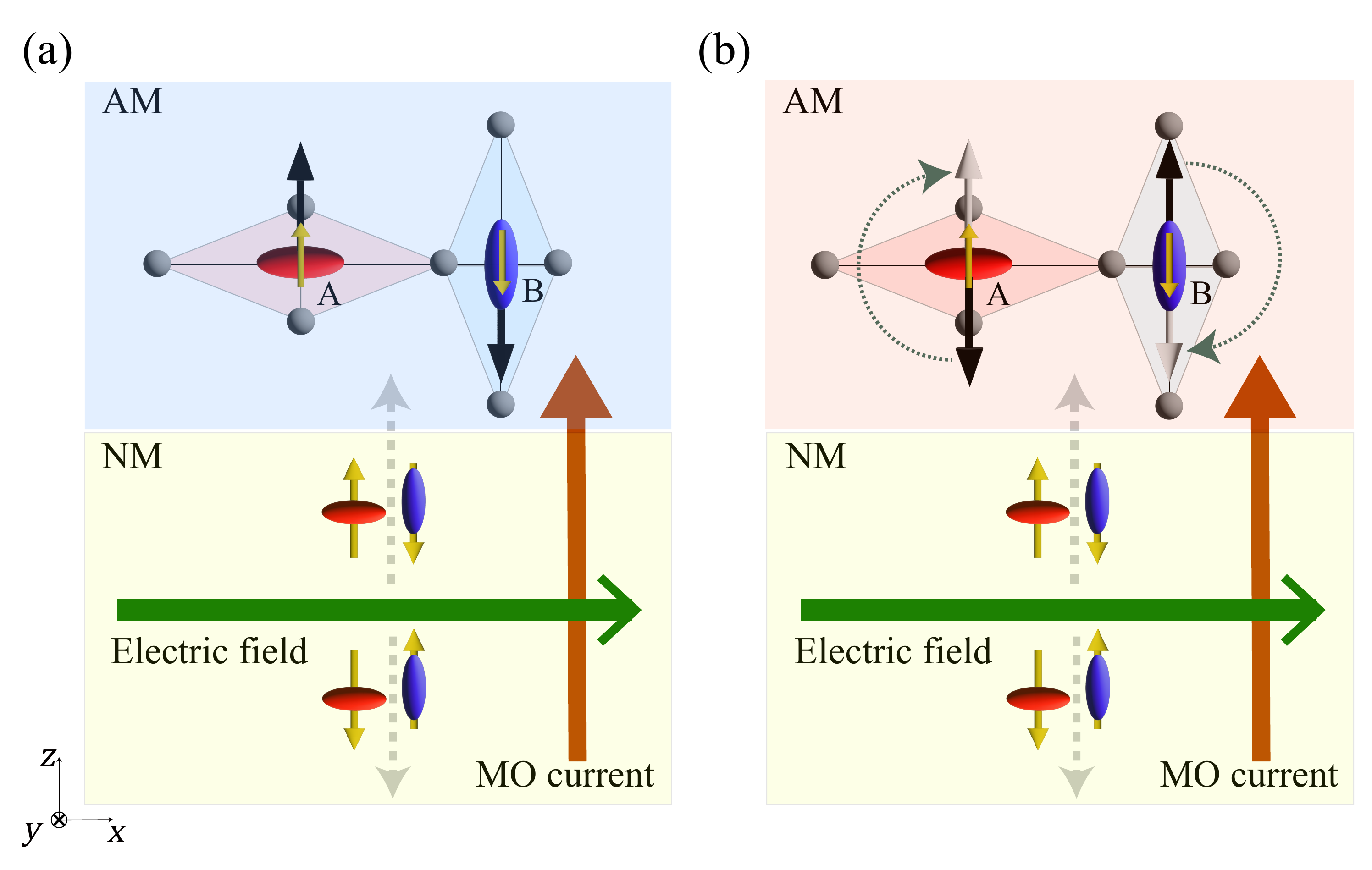}
  \subfigure{\label{fig1a}}
  \subfigure{\label{fig1b}} 
    \label{fig1}
\caption{  
Schematic of the N\'eel vector switching in an AM/NM bilayer for ${\bf N}$ along the $z$-direction (a) and $-z$-direction (b). An in-plane electric field (green arrows) along the $x$-direction generates an MO current (brown arrows) in the NM through the MO Hall effect~\cite{baek2025} and injects it to the AM. The yellow arrows in the NM denote the spin magnetic moment direction of conduction electrons, and the black arrows in the AM denote the magnetization direction of the local magnetic moment at the sublattices A and B of the AM. The horizontal (red) and vertical (blue) ellipses attached to the yellow and black arrows represent the orbital elongation directions. The yellow arrows in the AM denote the effective field generated by the injected MO. When the field is parallel (a) [anti-parallel (b)] to the magnetization direction of the local magnetic moments, the MO current does not [does] switch ${\bf N}$.
}
  \label{fig1}
\end{figure}
We assume that MO is externally injected into a $d$-wave AM. This can be achieved in an AM/normal metal (NM) bilayer (Fig.~\ref{fig1}) by applying an in-plane electric field to the NM so that the MO current generated by the MO Hall effect in the NM is injected into the AM (similar to the in-plane-current-induced spin current injection in the spin-orbit torque (SOT) geometry). It was recently predicted~\cite{baek2025} that many transition metals exhibit MO Hall effect with the MO Hall conductivity comparable to the spin Hall conductivities of Pt, Ta, and W. The NM side in Fig.~\ref{fig1a} illustrate schematically what an MO current (brown arrow) is microscopically: the electron velocity of conduction electrons (grey dashed arrows) is reversed when either electron spin magnetic moment direction (yellow solid arrows) or the electron orbital elongation direction [horizontal (red) vs vertical (blue) elongation] is flipped. When they are injected into an AM, they interact preferentially with those local moments (black solid arrows) that share the same orbital elongation direction (horizontal vs vertical). This orbital-elongation-direction dependence of the interaction becomes sublattice-selective interaction since the orbital elongation direction is different between the sublattices A and B. The sublattice-selective interaction then results in a staggered effective magnetic field along the MO electron spin magnetic moment directions in the sublattices A and B. The resulting field-like MOT on the N\'eel vector ${\bf N}$ of an AM is given by,
\begin{align}
    {\bf T}^{\text{MOT}} =  - {\bf N} \times \left(\sum_{i}g_{i} {\bf O}_{i}\right), \label{eq:T^MOT}
\end{align} 
which may be regarded as a dynamical implication of the coupling eneregy $-{\bf N}\cdot (\sum_i g_i {\bf O}_i)$ in equilibrium~\cite{mcclarty2024}. Here ${\bf O}_{i}$ is the MO density in an AM induced by the injected MO current injection. The vector direction of ${\bf O}_{i}$ denotes the spin magnetic moment polarization axis and the subscript $i$ specifies the electron orbital elongation direction of the injected MO electrons. On the other hand, the coupling constant $g_i$ describes the orbital elongation direction of the local moments. In the schematic in Fig.~\ref{fig1}, $g_i$ is nonzero only for $i=x^2 - z^2$, since $x^2 - z^2$ distinguishes the horizontal and vertical orbital elongation directions. In a rutile-type $d$-wave AM film, only $g_{xy}$ is non-zero for the film grown along the [001] direction, only $g_{x^2-z^2}$ is non-zero in the film along the [110] direction, and both $g_{xy}$ and $g_{yz}$ are non-zero for the film along the [101] direction~\cite{mcclarty2024,footnote1}. ${\bf T}^{\rm MOT}$ in Eq.~(\ref{eq:T^MOT}) acts analogously to a uniform field-like SOT that drives the N\'eel vector switching in collinear AFMs with hidden Rashba coupling~\cite{zelezny2014, bodnar2018}. However, it differs in two key aspects: (1) The MOT does not require the hidden Rashba coupling in AMs. (2) Instead, the MOT necessitates the MO ordering, and does not function in conventional collinear AFMs with $g_i=0$ for all $i$. In other words, the MOT is a unique intrinsic feature of $d$-wave AMs. By the way, we ignore the damping-like MOT in the subsequent discussion since its effective field is not staggered in general and not crucial for deterministic switching of ${\bf N}$.

The schematic in Fig.~\ref{fig1} allows one to guess the MOT-induced N\'eel vector dynamics in an AM. For simplicity, we assume only one $g_{i}$ is nonzero and positive, and consider the case when $\mathbf{O}_{i}$ for that $i$ is either parallel or antiparallel to the easy axis $\hat{\bf n}$ of ${\bf N}$. Then, ${\bf T}^{\text{MOT}}$ is expected to set ${\bf N}$ along the effective field direction $\mathbf{O}_{i}$. Thus, if ${\bf N}$ is initially antiparallel to $\mathbf{O}_{i}$, ${\bf T}^{\text{MOT}}$ will switch ${\bf N}$ to make it parallel to $\mathbf{O}_{i}$ [Fig.~\ref{fig1b}]. 
On the other hand, if ${\bf N}$ is initially parallel to $\mathbf{O}_{i}$, ${\bf T}^{\text{MOT}}$ will not switch ${\bf N}$ [Fig.~\ref{fig1a}]. As a result, the AM in a multi-domain magnetic configuration initially will be converted to a single domain. Below, we check this expectation through quantitative calculations.

\begin{figure}[!t]  
  \centering
  \includegraphics[width=8cm]{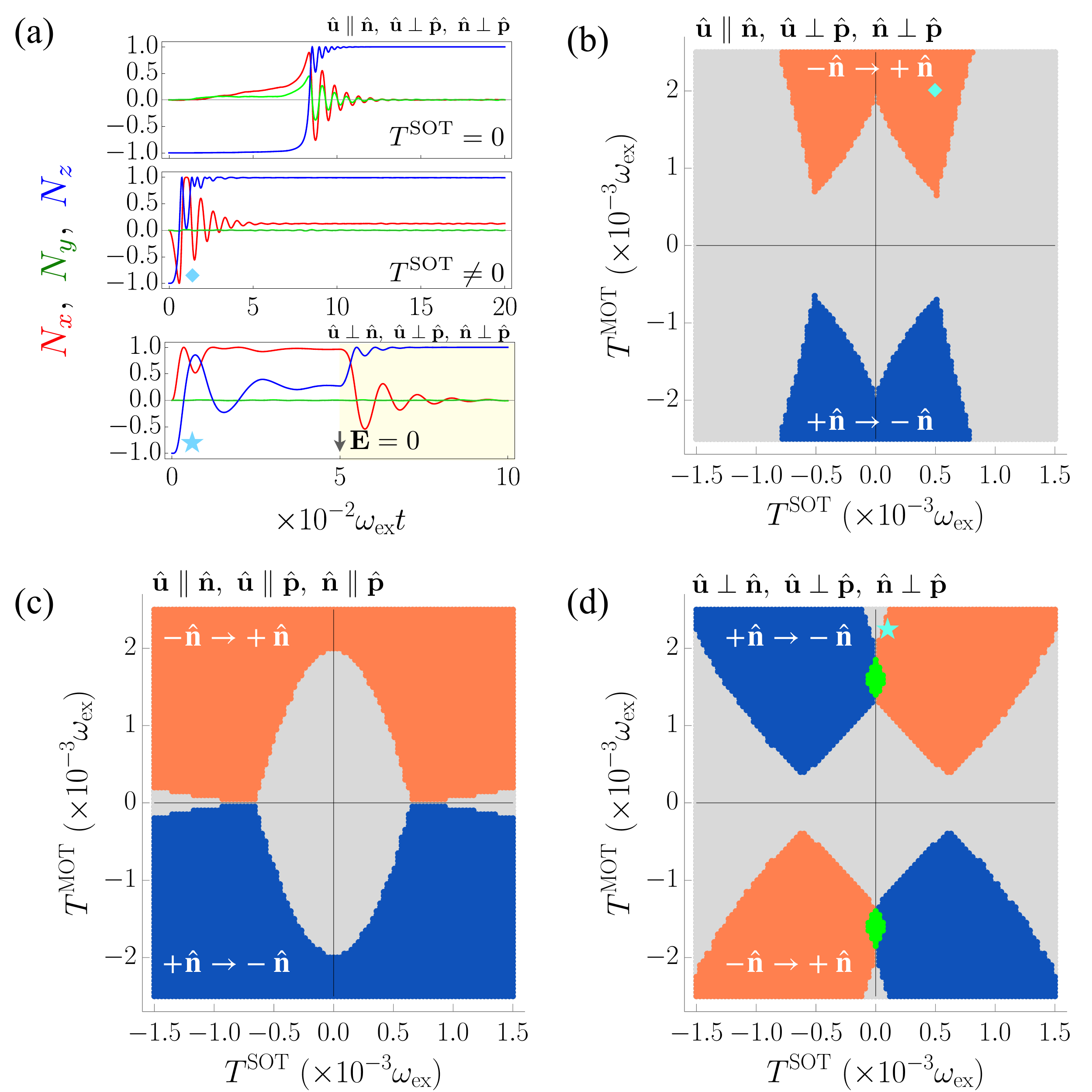}  
  \subfigure{\label{figMOTparallelDynamics}}
  \subfigure{\label{figMOTparallel}} 
  \subfigure{\label{figSOTMOTparallel}} 
  \subfigure{\label{figMOTperp}} 
  \caption 
   {
   (a) Top panel: The N\'eel vector dynamics induced by the MOT with $T^{\rm SOT}=0$ when the initial direction of ${\bf N}$ is along $\hat{\bf n}=-\hat{\bf z}$. Here, the MOT strength $T^{\rm MOT}$ is set to $2 \times 10^{-3} \omega_{\text{ex}}$. Middle (bottom) panel: The N\'eel vector dynamics induced jointly by the MOT and the SOT with $\hat{\bf u} \parallel \hat{\bf n} \perp \hat{\bf p}$ (with $\hat{\bf u}$, $\hat{\bf n}$, $\hat{\bf p}$ all perpendicular to each other). $T^{\rm MOT}$ and $T^{\rm SOT}$ are set to $2 \times 10^{-3} \omega_{\text{ex}}$ and $0.5\times 10^{-3} \omega_{\text{ex}}$ for the diamond symbol in (b), and $2.3 \times 10^{-3} \omega_{\text{ex}}$ and $0.1\times 10^{-3} \omega_{\text{ex}}$ for the star symbol in (d), respectively. In the bottom panel, the yellow transparent region shows the dynamics when both the MOT and the SOT are turned off. (b)--(d) Switching phase diagram for the $180^\circ$ switching of ${\bf N}$ when $\hat{\bf n}\parallel \hat{\bf u} \perp \hat{\bf p}$ (b), $\hat{\bf n}\parallel \hat{\bf u} \parallel \hat{\bf p}$ (c), and $\hat{\bf n} \perp \hat{\bf u} \perp \hat{\bf p} \perp \hat{\bf n}$ (d). In the orange-colored  regions, the switching occurs from $-\hat{\bf n}$ to $+\hat{\bf n}$, whereas in the dark blue-colored regions, the switching occurs the other way around. The deterministic switching is not possible in the grey-colored regions. The toggle switching~\cite{supple} occurs in the narrow green-colored region in (d).
   }
\label{fig02}
\end{figure} 

{\it 
Uniform domain dynamics--}
We examine quantitatively how each domain in an AM responds to the MOT injection. We take the spin Hamiltonian of an AM as follows,
\begin{equation}\label{spinHamiltonian}
    H_{\text{spin}} = \hbar \omega_{\text{ex}} {\bf S}_{\text{A}} \cdot {\bf S}_{\text{B}}
    - \sum_{\sigma=\text{A, B}}  \frac{\hbar \omega_{\text{easy}}}{2 S_{\sigma}^2}  ({\bf S}_{\sigma} \cdot \hat{\bf n} )^2   ,
\end{equation}
where ${\bf S}_{\sigma}$ ($\sigma=$ A, B) is the unit vector of the sublattice magnetization, $\hbar \omega_{\text{ex}}$ is the exchange interaction, $\hbar \omega_{\text{easy}}$ the magnetic easy-axis anisotropy. The possible anisotropy of exchange stiffness~\cite{gomonay2024} is neglected due to its irrelevance in single-domain AMs. The sublattice magnetization dynamics can be described by the Landau-Lifshitz-Gilbert equation, 
\begin{equation}\label{Eq:LLG}
    \dot{\bf S}_{\sigma} = - {\bf S}_{\sigma} \times {\bf H}_{\sigma}^{\text{eff}} 
    + \alpha {\bf S}_{\sigma} \times \dot{\bf S}_{\sigma} 
    + {\bf T}_{\sigma}^{\text{SOT}} 
    + {\bf T}_{\sigma}^{\text{MOT}} 
    + {\bf T}_{\sigma}^{\text{th}},
\end{equation}
where ${\bf H}^{\text{eff}}_{\sigma} = - \partial H_{\text{spin}} / \hbar \partial {\bf S}_{\sigma}$ is the effective magnetic field acting on sublattice $\sigma$ ($=$ A, B), and $\alpha$ is the Gilbert damping parameter. We consider both SOT and MOT since an in-plane electric field applied to an NM generates both of them. For the SOT, we consider only the damping-like SOT since its effective field is staggered, whereas the effective field for the field-like SOT is not. At the sublattice $\sigma$, the SOT and the MOT are given by ${\bf T}_{\sigma}^{\text{SOT}} = T^{\text{SOT}} {\bf S}_{\sigma} \times  ( {\bf S}_{\sigma} \times \hat{\bf p})$, and ${\bf T}_{\sigma}^{\text{MOT}} = - {\bf S}_{\sigma} \times \eta_\sigma (\sum_{i} g_{i} {\bf O}_{i}$)~\cite{han2024octupole}, where $\hat{\bf p}$ represents the spin polarization direction of the spin current injected into an AM and $\eta_{\text{A/B}} = \pm 1$. For simplicity, we assume only one component of $g_{i}$ is nonzero and set $\sum_{i} g_{i} \mathbf{O}_{i} = T^{\text{MOT}} \hat{\mathbf{u}}$, where $\hat{\mathbf{u}}$ is the unit vector of MO. Furthermore, we incorporate thermal fluctuations as ${\bf T}^{\text{th}}_{\sigma} = - {\bf S}_{\sigma} \times {\bf H}_{\sigma}^{\text{th}}$~\cite{akimoto2005,xu2023}, with the random field defined by ${\bf H}_{\sigma}^{\text{th}} = \mathcal{N}^{x}_{\sigma} (0, u) \hat{\bf x} + \mathcal{N}^{y}_{\sigma} (0, u) \hat{\bf y} + \mathcal{N}^{z}_{\sigma} (0, u) \hat{\bf z}$. Here, $\mathcal{N}_{\sigma}^{i=x, y, z}(0, u)$ denotes a normal distribution with zero mean and a standard deviation $u=\sqrt{2 \gamma \alpha k_{\text{B}} T / V_{\text{AM}} M_s (1+\alpha^2) \Delta t }$~\cite{akimoto2005,xu2023}, with $k_{\text{B}}$ denoting the Boltzmann constant, $\gamma$ the gyromagnetic ratio, $V_{\text{AM}} = 1000 \times 1000 \times 50$ nm$^3$ the volume of the altermagnet layer, $M_s = M_{s, \text{A}} = M_{s, \text{B}} = 1.5 \times 10^5$ A/m the saturation magnetization, $\Delta t = 1$ ps the fluctuation time interval, and $T = 300$ K the temperature. 

Using the Landau-Lifshitz-Gilbert equation [Eq.~\eqref{Eq:LLG}], we numerically calculate the response of ${\bf N}=({\bf S}_{\rm A}-{\bf S}_{\rm B} )/2$ to the MOT and the SOT. We first consider the case where $T^{\rm SOT}=0$ and only $T^{\rm MOT}$ is activated at time $t=0$ with $\hat{\bf u}=\hat{\bf n}$ and $T^{\rm MOT}>0$. We find that if ${\bf N}(t=0)=-\hat{\bf n}$, the MOT reverses ${\bf N}$ to $+\hat{\bf n}$, provided $T^{\rm MOT}$ is larger than a critical value [top panel in Fig.~\ref{figMOTparallelDynamics}]. On the other hand, if ${\bf N}(t=0)= + \hat{\bf n}$, the MOT leaves ${\bf N}$ essentially unchanged (not shown). This confirms the schematic in Fig.~\ref{fig1}. Before ${\bf N}$ is reversed in the top panel, a long incubation time appears. This is due to the fact that the MOT vanishes when ${\bf N}$ is perfectly anti-aligned with $\hat{\bf u}$. Once the thermal fluctuations tilt ${\bf N}$ sufficiently away from $-\hat{\bf u}$, the MOT reverses ${\bf N}$ quickly. This behavior is analogous to the magnetization reversal in an FM induced by the STT~\cite{miron2011, liu2012, manchon2019}.

When $T^{\rm SOT}$ is also activated at $t=0$ together with $T^{\rm MOT}$ ($\hat{\bf u}=\hat{\bf n}$), it may modify the MOT-induced dynamics. Figures~\ref{figMOTparallel} and \ref{figSOTMOTparallel} show the switching phase diagrams for this case with the spin polarization $\hat{\bf p}$ of the SOT perpendicular to and (anti)parallel with $\hat{\bf n}$, respectively. In the orange-colored regions, the joint application of $T^{\rm MOT}$ and $T^{\rm SOT}$ achieves the deterministic switching of ${\bf N}$ if ${\bf N}(t=0)=-\hat{\bf n}$ and does not switch ${\bf N}$ if ${\bf N}(t=0)=\hat{\bf n}$. In the dark blue-colored regions, the deterministic switching occurs the other way around. In the grey-colored regions, the deterministic switching does not occur. Regardless of the direction of $\hat{\bf p}$, we find that a moderate strength of $T^{\rm SOT}$ assists the MOT-driven switching and reduces the critical strength of $T^{\rm MOT}$ for switching. In particular, when $\hat{\bf p}$ is perpendicular to $\hat{\bf n}$, $T^{\rm SOT}$ also shortens the incubation time since the SOT tilts ${\bf N}$ away from the easy axis $\pm\hat{\mathbf{n}}$, thereby making the MOT more effective. Compared to the case with $T^{\text{SOT}}=0$ [top panel in Fig.~\ref{figMOTparallelDynamics}], the case with $T^{\text{SOT}}=0.5 \times 10^{-3}\omega_{\text{ex}}$ reduces the critical MOT strength by 62\% and the switching time by 84\% [middle panel in Fig.~\ref{figMOTparallelDynamics}]~\cite{footnote2}. On the other hand, very strong $T^{\rm SOT}$ makes the deterministric switching difficult and increases the critical MOT strength regardless of the direction of $\hat{\bf p}$ since the strong SOT leads to precessional dynamics and probabilistic switching.

Next, we consider the case where the MOT $\hat{\bf u}$ is perpendicular to $\hat{\bf n}$. From Eq.~(\ref{Eq:LLG}), we find that the MOT alone can achieve the probabilistic switching but the deterministic switching is not possible except for a narrow range of $T^{\rm MOT}$ [green-colored region in Fig.~\ref{fig02}(d)], where a toggle switching occurs, similarly to the SOT-induced toggle switching in FMs~\cite{toggle}. Details of toggle switching are presented in Ref.~\cite{supple}. Interestingly, we find that deterministic switching can be restored when the SOT is applied together with the MOT in a wide range of the phase diagram [Fig.~\ref{figMOTperp}] if the spin polarization $\hat{\bf p}$ of the SOT  is perpendicular to both $\hat{\bf u}$ and $\hat{\bf n}$. Thus, although either the MOT or the SOT alone can achieve only the probabilistic switching, their joint action sets a preferred direction of the switching and converts the probabilistic switching into the deterministic switching.

The deterministric switching in this case is analogous to the SOT-induced deterministic magnetization switching in FMs via the joint action of the SOT and an external magnetic field~\cite{manchon2019} when the SOT spin polarization, the field, and the easy axis of the magnetization are all perpendicular to each other. This analogy allows one to understand a few interesting differences between the deterministic switchings in the two cases: $\hat{\bf u}\perp\hat{\bf n}$ [Fig.~\ref{figMOTperp}] and $\hat{\bf u}\parallel \hat{\bf n}$ [Figs.~\ref{figMOTparallel} and \ref{figSOTMOTparallel}]. First, the switching dynamics in the former case is very different from the dynamics in the latter case. The joint action of $T^{\rm MOT}$ and $T^{\rm SOT}$ in the former case tilts ${\bf N}$ to a direction roughly perpendicular to $\hat{\bf n}$ and the deterministic switching is achieved only after $T^{\rm MOT}$ and $T^{\rm SOT}$ are turned off [bottom panel in Fig.~\ref{figMOTparallelDynamics}], whereas $T^{\rm MOT}$ in the latter case completes the deterministic switching while $T^{\rm MOT}$ is turned on [top and middle panels in Fig.~\ref{figMOTparallelDynamics}]. The former dynamics is analogous to the SOT-induced deterministic magnetization switching dynamics in FMs, whereas the latter dynamics to the STT-induced deterministic magnetization switching dynamics in FMs~\cite{manchon2019}. The second difference appears in the behavior of the ${\bf N}$ switching direction upon the in-plane electric field reversal. When an in-plane electric field is reversed, both $T^{\rm MOT}$ and $T^{\rm SOT}$ reverse their signs. Under this electric field reversal, the deterministic switching in the latter case ($\hat{\bf u}\parallel \hat{\bf n}$) reverses its switching direction, that is, the $(-\hat{\bf n})$-to-$(+\hat{\bf n})$ switching is reversed to the $(+\hat{\bf n})$-to-$(-\hat{\bf n})$ switching or vice versa [Figs.~\ref{figMOTparallel} and \ref{figSOTMOTparallel}]. In contrast, the in-plane electric field reversal in the former case ($\hat{\bf u}\perp\hat{\bf n}$) leaves the deterministic switching direction unchanged [Fig.~\ref{figMOTperp}]. This {\it unidirectional} switching in the former case can be understood by resorting to the analogy to the SOT-induced deterministic switching in FMs~\cite{manchon2019}: Since the deterministic switching direction in FMs is reversed by the SOT sign reversal and also by the external magnetic field reversal, the joint reversal of the SOT and the field leaves the deterministic switching direction in FMs unchanged.

Next, we discuss the relative directions of $\hat{\bf u}$, $\hat{\mathbf{p}}$, and $\hat{\mathbf{n}}$ for various AM/NM bilayer systems. When an in-plane electric field is applied to the bilayer along the $x(y)$-direction, the MO Hall effect injects into the AM various components of MO currents, including $O_{xy}^x$($O_{xy}^y$), $O_{yz}^z$($O_{xz}^z$), $O_{x^2}^y$($O_{x^2}^x$), $O_{y^2}^y$($O_{y^2}^x$), and $O_{z^2}^y$($O_{z^2}^x$)~\cite{baek2025}. Out of these components, only one (or at best two) component matches the orbital elongation direction of the equilibrium MO ordering in the AM and generates the MOT. On the other hand, the SHE injects into the AM spin currents with the spin polarization along the $y$($x$)-direction. In the FeSb$_2$/NM bilayer, the easy axis $\hat{\bf n}$ lies along the $y$-direction~\cite{mazin2021} with the equilibrium MO ordering of $O_{xy}^y$. Then, an electric field application along the $y$-direction generates the MO current of $O^y_{xy}$, thereby realizing the situation with $\hat{\bf n} \ (\hat{\bf y}) \parallel \hat{\bf u} \ (O^y_{xy}) \perp {\bf p} \ (\hat{\bf x})$ [Fig.~\ref{figMOTparallel}]. In the RuO$_2$[110]/NM bilayer, $\hat{\bf n}$ lies along the $y$-direction~\cite{mcclarty2024,schiff2024} with the equilibrium MO ordering of $O_{x^2-z^2}^y$. Then, an electric field along the $x$-direction generates the MO current of $O^y_{x^2-y^2}$, thereby realizing the situation with $\hat{\bf n} \ (\hat{\bf y}) \parallel \hat{\bf u} \ (O^y_{x^2-y^2}) \parallel {\bf p} \ (\hat{\bf y})$ [Fig.~\ref{figSOTMOTparallel}]. In the RuO$_2$[001]/NM or MnF$_2$/NM bilayer, $\hat{\bf n}$ is typically along the $z$-direction~\cite{bhowal2024,mcclarty2024,schiff2024} with the equilibrium MO ordering of $O_{xy}^z$. Then, an electric field along the $x$-direction realizes the situation with $\hat{\bf n} \ (\hat{\bf z})$, $\hat{\bf u} \ (O^x_{xy})$, $\hat{{\bf p}} \ (\hat{\bf y})$ all perpendicular to each other [Fig.~\ref{figMOTperp}]. Therefore, the deterministic N\'eel vector switching is possible in all these bilayers.

\begin{figure}[]
  \centering
  \subfigure{\label{fig3a}}
  \subfigure{\label{fig3b}} 
  \subfigure{\label{fig3c}} 
  \includegraphics[width=8.8cm]{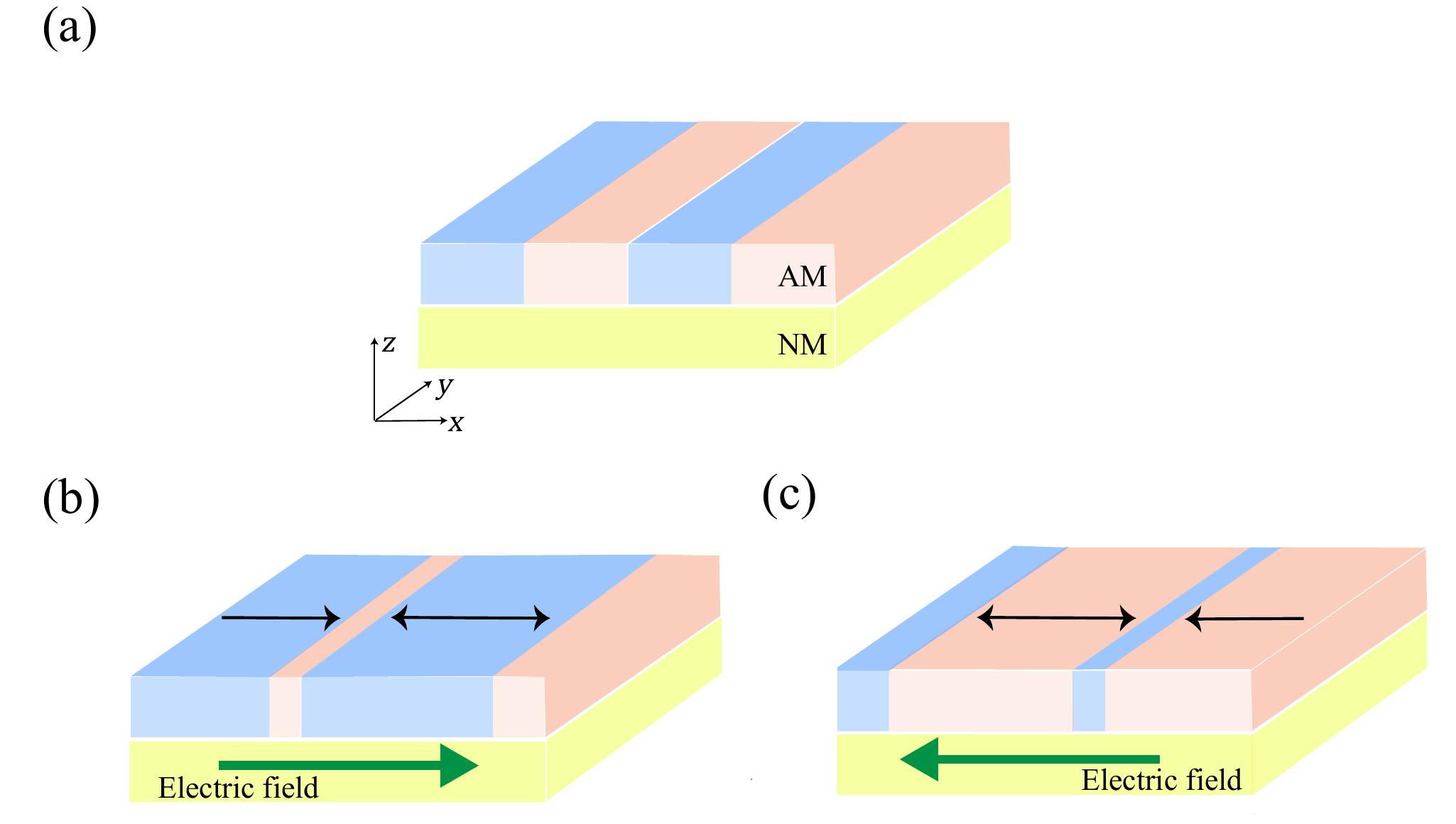} 
\caption 
{Schematics of the DW dynamics in AMs driven bythe MOT in AM/NM bilayers.
(a) Schematic of an AM/NM bilayer 
with the AM in a multidomain configuration.
Light blue and light pink regions represent domains with N\'eel vectors aligned along $\hat{\mathbf{n}}$ and $-\hat{\mathbf{n}}$, respectively. (b), (c) When an electric field (green arrow) is applied, the resulting MOT induces the DW motion. For $\hat{\bf u} \parallel \hat{\bf n}$, positive $T^{\rm MOT}_\parallel$ drives the light blue domains to expand and the light pink domains to shrink (b). Negative $T^{\rm MOT}_\parallel$ drives the domain expansion and shrinkage the other way around (c).} 

  \label{fig3}
\end{figure}

{\it Domain wall dynamics}-- The domain wall (DW) motion is an important element of the magnetization dynamics. Here, we examine the DW motion in AMs induced by the application of the MOT and/or the SOT. For this study, we use the collective coordinate approach~\cite{tveten2013,shiino2016}, which describes the DW dynamics in terms of the DW center position $r(t)$ and the DW center angle $\phi(t)$. In the presence of the MOT and the SOT, the time evolution of the two collective coordinates are given by 
%
%
\begin{align} 
\ddot{r}+a\gamma l\alpha \dot{r}\pm\frac{\pi}{2}a\gamma^2 l\lambda T^{\text{SOT}}_\perp\cos\phi  \mp a\lambda\gamma T^{\text{MOT}}_\parallel &=0, \label{DW_eq_r}\\
\ddot{\phi} +a\gamma l\alpha \dot{\phi}-a\gamma^2lT^{\text{SOT}}_\parallel +\frac{\pi}{2}a\gamma T^{\text{MOT}}_\perp \sin\phi&=0, \label{DW_eq_phi}
\end{align}
where $l$ represents magnitude of the N\'eel vector, $\gamma$ is the gyromagnetic ratio, $a$ denote the homogeneous exchange constants, and $\lambda$ is the domain wall width.
$T^{\rm MOT}_{\parallel}$ and $T^{\rm MOT}_{\perp}$ amount to the field-like MOT $T^{\rm MOT}$ for the cases with $\hat{\bf u} \parallel \hat{\bf n}$ and $\hat{\bf u} \perp \hat{\bf n}$, respectively. $T^{\rm SOT}_{\parallel}$ and $T^{\rm SOT}_{\perp}$ amount to the damping-like SOT $T^{\rm SOT}$ for the cases with $\hat{\bf p} \parallel \hat{\bf n}$ and $\hat{\bf p} \perp \hat{\bf n}$, respectively. The upper and lower signs in Eq.~\eqref{DW_eq_r} apply to the light blue-to-light pink DWs and the light pink-to-light blue DWs [Fig.~\ref{fig3a}], respectively. Here, the light blue and the light pink domains have ${\bf N}$ along the $+\hat{\bf n}$ and $-\hat{\bf n}$ directions, respectively. 


First, we consider the case with $T^{\rm MOT}_\parallel\neq 0$ and $T^{\rm SOT}=0$ with $\hat{\bf u}\parallel \hat{\bf n}$.
In this case, Eq.~\eqref{DW_eq_r} predicts the steady state DW velocity of $\dot{r}^{\rm steady}=\pm \lambda T^{\rm MOT}_\parallel/(\alpha l)$, which has opposite signs for the light blue-to-light pink and the light pink-to-light blue DWs. Thus, for a positive $T^{\rm MOT}_\parallel$, the light blue (light pink) domains expand (shrink) [Fig.~\ref{fig3b}]. For a negative $T^{\rm MOT}_\parallel$, it is the other way around [Fig.~\ref{fig3c}]. Therefore, the DW motion induced by $T^{\rm MOT}$ tends to convert a multi-domain configuration into a single domain [light blue (light pink) domain for positive (negative) $T^{\rm MOT}_\parallel$]. Note that the resulting single domain 
agrees with the 
single domain enforced by the uniform domain switching considered above [Fig.~\ref{figMOTparallel} and \ref{figSOTMOTparallel}]. The DW velocity does not depend on DW type (Bloch vs N\'eel).

Next, we consider the joint application of 
$T^{\rm MOT}_\perp\neq 0$ and $T^{\rm SOT}_\perp\neq 0$ with $\hat{\bf u}$, $\hat{\bf p}$, and $\hat{\bf n}$ all perpendicular to each other. In this case, Eq.~\eqref{DW_eq_r} predicts the steady state DW velocity of $\dot{r}^{\rm steady}=\mp (\pi/2) (\gamma \lambda/\alpha) T^{\rm SOT}_\perp \cos \phi^{\rm steady}$, where the steady state DW angle $\phi^{\rm steady}$ is determined by Eq.~\eqref{DW_eq_phi} to be $\cos\phi^{\rm steady}={\rm sgn}(T^{\rm MOT}_\perp)$. Here sgn($\cdots$) denotes the sign function. We then obtain $\dot{r}^{\rm steady}=\mp (\pi/2) (\gamma  \lambda/\alpha) T^{\rm SOT}_\perp {\rm sgn}(T^{\rm MOT}_\perp)$, which has opposite signs for the light blue-to-light pink and light pink-to-light blue DWs. Thus, for positive $T^{\rm MOT}_\perp T^{\rm SOT}_\perp$, the light pink (light blue) domains expand (shrhink) [similar to Fig.~\ref{fig3c}]. For negative $T^{\rm MOT}_\perp T^{\rm SOT}_\perp$, it is the other way around [similar to Fig.~\ref{fig3b}]. Therefore, the DW motion induced jointly by $T^{\rm MOT}_\perp$ and $T^{\rm SOT}_\perp$ tends to convert a multi-domain configuration into a single domain [light pink (light blue) domain for positive (negative) $T^{\rm MOT}_\perp T^{\rm SOT}_\perp$]. Note that the resulting single domain again agrees with the 
single domain enforced by 
the uniform domain switching in Fig.~\ref{figMOTperp}. Note also that an in-plane electric field reversal, which reverses the signs of both $T^{\rm MOT}_\perp$ and $T^{\rm SOT}_\perp$, does not reverse the DW motion direction, which is consistent with the unidirectional switching in Fig.~\ref{figMOTperp}. 
%

{\it Discussion--}
Our switching mechanism 
offers several key advantages. First, the MO Hall effect in NMs responsible for the MOT in AMs
does not require any symmetry breaking and thus can remain operative 
even in polycrystalline NM layers.
Second, our mechanism is stable with respect to at least moderate amount of the AM/NM interface roughness. For instance, in the A-type AFM ordering in AMs, when the interface layer on the AM side fluctuates between the sublattices A and B region by region, 
our switching mechanism remains active since the staggered effective magnetic field of the MOT arises from the orbital-elongation-direction dependence rather than the sublattice location.
Third, since the MO Hall conductivity and spin Hall conductivity vary across different NM materials~\cite{baek2025}, this variation provides an additional degree of freedom to engineer the electrical deterministic switching behavior of AMs.

\rev{Our switching mechanism applies 
also to $g$-wave AMs. Due to the coupling in $g$-wave AMs between their N\'eel vector and the rank-5 magnetic triakontadipole (MT)~\cite{mcclarty2024, verbeek2024}, the injection of the MT into $g$-wave AMs generates the MT torque, which generates the staggered effective fields on the sublattices of the AMs and allows for the deterministic switching of the N\'eel vector. The MT injection can be achieved in $g$-wave AM/NM bilayers through the MT Hall effect in NMs. For instance, the MT Hall effect is sizable in Pt~\cite{supple}. A recent experiment~\cite{amin2024} on $g$-wave AM/Pt bilayers reported field-free switching in $g$-wave AMs and attributed it to the joint action of the SOT and the Dzyaloshinskii-Moriya interaction. We speculate 
that the MT torque may be partially responsible for the reported switching.}

Our switching mechanism based on magnetic multipoles may be combined with other switching mechanisms~\cite{baek2018,ryu2022}. For an AM1/NM/AM2 trilayer, suppose a single domain configuration can be achieved in the AM1 by other switching mechanisms (magnetic field~\cite{amin2024} for instance). Then, an out-of-plane-direction electron flow from the AM1 to the AM2 can generate a magnetic multipole torque on the AM2~\cite{ko2025}, which can be utilized for the deterministic switching of the AM2. This switching resembles the switching in FM1/NM/FM2 trilayers caused by out-of-plane-current-induced STT~\cite{baek2018,ryu2022}.

An NM1/AM/NM2 trilayer may allow for additional functionalities. Suppose the in-plane current $I_1$ and $I_2$ to the NM1 and NM2 are controlled separately by a word line and bit line, respectively, and the NM1 and NM2 generates mainly the MOT and SOT, respectively, with their directions depicted in Fig.~\ref{figMOTperp}. Then, reversing the sign of $I_2$ reverses the switching behavior caused by I1, realizing a reconfigurable switching device.  

To conclude, we demonstrated that the N\'eel vector in 
AMs can be deterministically switched via magnetic multipole torque, 
without requiring any external magnetic field or Dzyaloshinskii-Moriya interaction. By harnessing magnetic multipoles as active control parameters, our study introduces a new paradigm for manipulating the N\'eel vector in AMs. This approach 
facilitates domain alignment, which is essential for realizing the intrinsic physical properties of AMs—making our findings of both technological and fundamental significance.

\begin{acknowledgments}
We thank Changyoung Kim, Sang-Wook Cheong, and Daegeun Jo for the fruitful discussions. This work was financially supported by the National Research Foundation of Korea (NRF) grant funded by the Korean government (MSIT) (No. RS-2024-00356270, RS-2024-00334933, RS-2024-00410027)
\end{acknowledgments}


\begin{thebibliography}{99}%

\bibitem{hayami2019}
S. Hayami, Y. Yanagi, and H. Kusunose,
Momentum-Dependent Spin Splitting by Collinear Antiferromagnetic Ordering,
JPSJ \textbf{88}, 123702 (2019).

\bibitem{naka2019}
M. Naka, S. Hayami, H. Kusunose, Y. Yanagi, Y. Motome, and H. Seo,
Spin current generation in organic antiferromagnets,
Nat. Commun. \textbf{10}, 12229 (2019).

\bibitem{ahn2019}
K.-H. Ahn, A. Hariki, K.-W. Lee, and J. Kuneš,
Antiferromagnetism in RuO$_2$ as $d$-wave Pomeranchuk instability,
Phys. Rev. B \textbf{99}, 184432 (2019).

\bibitem{yuan2020}
L.-D. Yuan, Z. Wang, J.-W. Luo, E. I. Rashba, and A. Zunger,
Giant momentum-dependent spin splitting in centrosymmetric low-$Z$ antiferromagnets,
Phys. Rev. B \textbf{102}, 014422 (2020).

\bibitem{igor2021}
I. I. Mazin, K. Koepernik, M. D. Johannes, R. González-Hernández, and L. Šmejkal,
Prediction of unconventional magnetism in doped FeSb$_2$,
PNAS \textbf{118}, e2108924118 (2021).

\bibitem{yuan2021}
L.-D. Yuan, Z. Wang, J.-W. Luo, and A. Zunger,
Prediction of low-$Z$ collinear and noncollinear antiferromagnetic compounds having momentum-dependent spin splitting even without spin–orbit coupling,
Phys. Rev. Mater. \textbf{5}, 014409 (2021).

\bibitem{yuan2022_1}
L.-D. Yuan and A. Zunger,
Degeneracy removal of spin bands in antiferromagnets with non-interconvertible spin-motif pair,
Adv. Mater. \textbf{35}, 2211966 (2023).

\bibitem{yuan2022_2}
L.-D. Yuan, X. Zhang, C. Mera, and A. Zunger,
Uncovering hidden spin polarization of energy bands in antiferromagnets,
Nat. Commun. \textbf{14}, 5301 (2023).

\bibitem{wu2007}
C. Wu, K. Sun, E. Fradkin, and S.-C. Zhang,
Fermi liquid instabilities in the spin channel,
Phys. Rev. B \textbf{75}, 115103 (2007).

\bibitem{noda2016}
Y. Noda, K. Ohno, and S. Nakamura,
Momentum-dependent band spin splitting in semiconducting MnO$_2$: a density-functional calculation,
Phys. Chem. Chem. Phys. \textbf{18}, 13294–13303 (2016).

\bibitem{ma2021}
H.-Y. Ma, M. Hu, N. Li, J. Liu, W. Yao, J.-F. Jia, and J. Liu,
Multifunctional antiferromagnetic materials with giant piezomagnetism and noncollinear spin current,
Nat. Commun. \textbf{12}, 2846 (2021).

\bibitem{smejkal2022_1}
L. Šmejkal, J. Sinova, and T. Jungwirth,
Emerging research landscape of altermagnetism,
Phys. Rev. X \textbf{12}, 040501 (2022).

\bibitem{smejkal2022_2}
L. Šmejkal, J. Sinova, and T. Jungwirth,
Beyond conventional ferromagnetism and antiferromagnetism: a phase with nonrelativistic spin and crystal-rotation symmetry,
Phys. Rev. X \textbf{12}, 031042 (2022).

\bibitem{lee2024}
S. Lee, S. Lee, S. Jung, J. Jung, D. Kim, Y. Lee, B. Seok, J. Kim, B. G. Park, L. Šmejkal, and T. Jungwirth,
Broken Kramers degeneracy in altermagnetic MnTe,
Phys. Rev. Lett. \textbf{132}, 036702 (2024).

\bibitem{han2024}
L. Han, X. Fu, R. Peng, X. Cheng, J. Dai, L. Liu, Y. Li, Y. Zhang, W. Zhu, H. Bai, Y. Zhou, S. Liang, C. Chen, Q. Wang, X. Chen, L. Yang, Y. Zhang, C. Song, J. Liu, and F. Pan,
Electrical 180$^\circ$ switching of Néel vector in spin-splitting antiferromagnet,
Sci. Adv. \textbf{10}, eadn0479 (2024).

\bibitem{shao2021}
D.-F. Shao, S.-H. Zhang, M. Li, C.-B. Eom, and E. Y. Tsymbal,
Spin-neutral currents for spintronics,
Nat. Commun. \textbf{12}, 26915 (2021).

\bibitem{smejkal2020}
L. Šmejkal, R. González-Hernández, T. Jungwirth, and J. Sinova,
Crystal time-reversal symmetry breaking and spontaneous Hall effect in collinear antiferromagnets,
Sci. Adv. \textbf{6}, eaaz8809 (2020).

\bibitem{gonzalez2021}
R. González-Hernández, L. Šmejkal, K. Výborný, Y. Yahagi, J. Sinova, T. Jungwirth, and J. Železný,
Efficient electrical spin splitter based on nonrelativistic collinear antiferromagnetism,
Phys. Rev. Lett. \textbf{126}, 127701 (2021).

\bibitem{bose2022}
A. Bose, N. J. Schreiber, R. Jain, D.-F. Shao, H. P. Nair, J. Sun, X. S. Zhang, D. A. Muller, E. Y. Tsymbal, D. G. Schlom, and D. C. Ralph,
Tilted spin current generated by the collinear antiferromagnet ruthenium dioxide,
Nat. Electron. \textbf{5}, 267 (2022).

\bibitem{karube2022}
S. Karube, T. Tanaka, D. Sugawara, N. Kadoguchi, M. Kohda, and J. Nitta,
Observation of spin-splitter torque in collinear antiferromagnetic RuO$_2$,
Phys. Rev. Lett. \textbf{129}, 137201 (2022).

\bibitem{liu2024}
F. Liu, Z. Zhang, X. Yuan, Y. Liu, S. Zhu, Z. Lu, and R. Xiong,
Giant tunneling magnetoresistance in insulated altermagnet/ferromagnet junctions induced by spin-dependent tunneling effect,
Phys. Rev. B \textbf{110}, 134437 (2024).

\bibitem{noh2025}
S. Noh, G.-H. Kim, J. Lee, H. Jung, U. Seo, G. So, J. Lee, S. Lee, M. Park, S. Yang,
Y. S. Oh, H. Jin, C. Sohn, and J.-W. Yoo,
Tunneling magnetoresistance in altermagnetic RuO$_2$-based magnetic tunnel junctions,
Phys. Rev. Lett. \textbf{134}, 246703 (2025).

\bibitem{zhou2025}
Z. Zhou, X. Cheng, M. Hu, R. Chu, H. Bai, L. Han, J. Liu, F. Pan, and C. Song,
Manipulation of the altermagnetic order in CrSb via crystal symmetry,
Nature \textbf{638}, 645–650 (2025).

\bibitem{amin2024}
O. J. Amin, A. Dal Din, E. Golias, Y. Niu, A. Zakharov, S. C. Fromage, C. J. B. Fields, S. L. Heywood, R. B. Cousins, J. Krempaský, J. H. Dil, D. Kriegner, B. Kiraly, R. P. Campion, A. W. Rushforth, K. W. Edmonds, S. S. Dhesi, L. Šmejkal, T. Jungwirth, and P. Wadley,
Nanoscale imaging and control of altermagnetism in MnTe,
Nature \textbf{636}, 348–353 (2024).

\bibitem{chen2025}
Y. Chen, X. Liu, H.-Z. Lu, and X. C. Xie,
Electrical switching of altermagnetism,
Phys. Rev. Lett. \textbf{135}, 016701 (2025).

\bibitem{song2025}
Q. Song, S. Stavrić, P. Barone, A. Droghetti, D. S. Antonenko, J. W. F. Venderbos, C. A. Occhialini, B. Ilyas, E. Ergeçen, N. Gedik, S.-W. Cheong, R. M. Fernandes, S. Picozzi, and R. Comin,
Electrical switching of a p-wave magnet,
Nature \textbf{642}, 64--70 (2025).

\bibitem{shao2023}
D.-F. Shao, Y.-Y. Jiang, J. Ding, S.-H. Zhang, Z.-A. Wang, R.-C. Xiao, G. Gurung, W. J. Lu, Y. P. Sun, and E. Y. Tsymbal,
Néel spin currents in antiferromagnets,
Phys. Rev. Lett. \textbf{130}, 216702 (2023).

\bibitem{zhang2025}
S.-S. Zhang, Z.-A. Wang, B. Li, W.-J. Lu, M. Tian, Y.-P. Sun, H. Du, and D.-F. Shao,
Deterministic switching of the Néel vector by asymmetric spin torque,
arXiv:2506.10786 (2025).

\bibitem{han2024octupole}
S. Han, D. Jo, I. Baek, P. M. Oppeneer, and H.-W. Lee,
Harnessing magnetic octupole Hall effect to induce torque in altermagnets,
arXiv:2409.14423 (2024).

\bibitem{bhowal2024}
S. Bhowal and N. A. Spaldin,
Ferroically ordered magnetic octupoles in $d$-wave altermagnets,
Phys. Rev. X \textbf{14}, 011019 (2024).

\bibitem{mcclarty2024}
P. A. McClarty and J. G. Rau,
Landau theory of altermagnetism,
Phys. Rev. Lett. \textbf{132}, 176702 (2024).

\bibitem{verbeek2024}
X. H. Verbeek, D. Voderholzer, S. Schären, Y. Gachnang, N. A. Spaldin, and S. Bhowal,
Non-relativistic ferromagnetotriakontadipolar order and spin splitting in hematite,
Phys. Rev. Res. \textbf{6}, 043157 (2024).

\bibitem{khvalkovskiy2013}
A. V. Khvalkovskiy, V. Cros, D. Apalkov, V. Nikitin, M. Krounbi, K. A. Zvezdin, A. Anane, J. Grollier, and A. Fert,
Matching domain-wall configuration and spin-orbit torques for efficient domain-wall motion,
Phys. Rev. B \textbf{87}, 020402 (2013).

\bibitem{gambardella2011}
P. Gambardella and I. M. Miron,
Current-induced spin–orbit torques,
Philos. Trans. R. Soc. A \textbf{369}, 3175–3197 (2011).

\bibitem{manchon2019}
A. Manchon, J. Železný, I. M. Miron, T. Jungwirth, J. Sinova, A. Thiaville, K. Garello, and P. Gambardella,
Current-induced spin-orbit torques in ferromagnetic and antiferromagnetic systems,
Rev. Mod. Phys. \textbf{91}, 035004 (2019).

\bibitem{baek2025}
I. Baek, S. Han, and H.-W. Lee,
Magnetic octupole Hall effect in heavy transition metals,
arXiv:2507.02516 (2025).

\bibitem{footnote1}
In general, ${\bf O}_{i}$ consists of both equilibrium and externally injected contributions. Here, we consider only the externally injected contribution, since the equilibrium ${\bf O}_{i}$ is aligned parallel to the N\'eel vector.

\bibitem{zelezny2014}
J. Železný, H. Gao, K. Výborný, J. Zemen, J. Mašek, A. Manchon, J. Wunderlich, J. Sinova, and T. Jungwirth,
Relativistic Néel-order fields induced by electrical current in antiferromagnets,
Phys. Rev. Lett. \textbf{113}, 157201 (2014).

\bibitem{bodnar2018}
S. Yu. Bodnar, L. Šmejkal, I. Turek, T. Jungwirth, O. Gomonay, J. Sinova, A. A. Sapozhnik, H.-J. Elmers, M. Kläui, and M. Jourdan,
Writing and reading antiferromagnetic Mn$_2$Au by Néel spin–orbit torques and large anisotropic magnetoresistance,
Nat. Commun. \textbf{9}, 348 (2018).

\bibitem{gomonay2024}
O. Gomonay, V. P. Kravchuk, R. Jaeschke-Ubiergo, K. V. Yershov, T. Jungwirth, L. Šmejkal, J. van den Brink, and J. Sinova,
Structure, control, and dynamics of altermagnetic textures,
npj Spintronics \textbf{2}, 35 (2024).

\bibitem{xu2023}
Z. Xu, J. Ren, Z. Yuan, Y. Xin, X. Zhang, S. Shi, Y. Yang, and Z. Zhu,
Field-free spin–orbit-torque switching of an antiferromagnet with perpendicular Néel vector,
J. Appl. Phys. \textbf{133}, 153904 (2023).

\bibitem{akimoto2005}
H. Akimoto, H. Kanai, Y. Uehara, T. Ishizuka, and S. Kameyama,
Analysis of thermal magnetic noise in spin-valve GMR heads by using micromagnetic simulation,
J. Appl. Phys. \textbf{97}, 10N705 (2005).

\bibitem{miron2011}
I. M. Miron, K. Garello, G. Gaudin, P.-J. Zermatten, M. V. Costache, S. Auffret, S. Bandiera, B. Rodmacq, A. Schuhl, and P. Gambardella,
Perpendicular switching of a single ferromagnetic layer induced by in-plane current injection,
Nature \textbf{476}, 189--193 (2011).

\bibitem{liu2012}
L. Liu, C.-F. Pai, Y. Li, H. W. Tseng, D. C. Ralph, and R. A. Buhrman,
Spin-torque switching with the giant spin Hall effect of tantalum,
Science \textbf{336}, 555--558 (2012).

\bibitem{footnote2}
From 30 independent trials considering thermal fluctuations, the obtained average switching time is $9.66 \pm 2.27$ without SOT and $1.50 \pm 0.047$ with SOT, in unit of $(10^{-2} \omega_{\text{ex}})^{-1}$

\bibitem{toggle}
N. Hassan, S. P. Lainez-Garcia, F. Garcia-Sanchez, and J. S. Friedman,
Toggle Spin-Orbit Torque MRAM with perpendicular Magnetic Anisotropy,  IEEE J. Exploratory Solid-State
Comput. Devices Circuits 5, 166 (2019).

\bibitem{supple} See the Supplemental Materials.

\bibitem{mazin2021}
I. I. Mazin, K. Koepernik, M. D. Johannes, R. González-Hernández, and L. Šmejkal,
Prediction of unconventional magnetism in doped FeSb$_2$,
PNAS \textbf{118}, e2108924118 (2021).

\bibitem{schiff2024}
H. Schiff, P. McClarty, J. G. Rau, and J. Romhányi,
Collinear altermagnets and their Landau theories,
arXiv:2412.18025 (2024).

\bibitem{tveten2013}
E. G. Tveten, A. Qaiumzadeh, O. A. Tretiakov, and A. Brataas,
Staggered dynamics in antiferromagnets by collective coordinates,
Phys. Rev. Lett. \textbf{110}, 127208 (2013).

\bibitem{shiino2016}
T. Shiino, S.-H. Oh, P. M. Haney, S.-W. Lee, G. Go, B.-G. Park, and K.-J. Lee,
Antiferromagnetic domain-wall motion driven by spin–orbit torques,
Phys. Rev. Lett. \textbf{117}, 087203 (2016).

\bibitem{baek2018}
S.-H. C. Baek, V. P. Amin, Y.-W. Oh, G. Go, S.-J. Lee, G.-H. Lee, K.-J. Kim, M. D. Stiles, B.-G. Park, and K.-J. Lee,
Spin currents and spin–orbit torques in ferromagnetic trilayers,
Nat. Mater. \textbf{17}, 509--513 (2018).

\bibitem{ryu2022}
J. Ryu, R. Thompson, J. Y. Park, S.-J. Kim, G. Choi, J. Kang, H. B. Jeong, M. Kohda, J. M. Yuk, J. Nitta, K.-J. Lee, and B.-G. Park,
Efficient spin–orbit torque in magnetic trilayers using all three polarizations of a spin current,
Nat. Electron. \textbf{5}, 217--223 (2022).

\bibitem{ko2025}
H.-W. Ko and K.-J. Lee,
Magnetic octupole Hall effect in d-wave altermagnets,
arXiv:2508.00794 (2025).


















\end{thebibliography}
\end{document}